\documentclass[12pt,letter,useAMS]{emulateapj-rtx4}
\usepackage{graphicx,amssymb}
\usepackage{natbib,txfonts}
\slugcomment{draft}

\bibpunct[, ]{(}{)}{;}{a}{}{,}
\def \aj {AJ}
\def \mnras {MNRAS}
\def \pasp {PASP}
\def \apj {ApJ}
\def \apjs {ApJS}
\def \apjl {ApJL}
\def \aap {A\&A}
\def \nat {Nature}
\def \araa {ARAA}
\def \iaucirc {IAUC}
\def \amp {\&}
\newcommand{\kms} {$\mathrm{ km \; s^{-1}}\,$}

\def\lesssim{\mathrel{\hbox{\rlap{\hbox{\lower4pt\hbox{$\sim$}}}\hbox{$<$}}}}
\def\gtrsim{\mathrel{\hbox{\rlap{\hbox{\lower4pt\hbox{$\sim$}}}\hbox{$>$}}}}

\long\def\symbolfootnote[#1]#2{\begingroup%
\def\thefootnote{\fnsymbol{footnote}}\footnote[#1]{#2}\endgroup}

\begin{document}

\title{A spectropolarimetric view on the nature of the peculiar Type I SN~2005hk\footnote{Based on observations made with ESO Telescopes at the Paranal Observatory, under program 076.D-0178(A).}}

\author{Justyn R. Maund\altaffilmark{1,2,3},
  J.~Craig~Wheeler\altaffilmark{4}, Lifan Wang\altaffilmark{5},
  Dietrich Baade\altaffilmark{6}, Alejandro
  Clocchiatti\altaffilmark{7},\\Ferdinando Patat\altaffilmark{6},
  Peter H\"oflich\altaffilmark{8}, Jason Quinn\altaffilmark{7}, and Paula Zelaya\altaffilmark{7}}
\altaffiltext{1}{Dark Cosmology Centre, Niels Bohr Institute,
  University of Copenhagen, Juliane Maries Vej, DK-2100 Copenhagen \O,
  Denmark; justyn@dark-cosmology.dk} 

\altaffiltext{2}{Department of Astronomy \& Astrophysics, University of California, Santa Cruz, 95064, U.S.A.}

\altaffiltext{3}{Sophie \& Tycho Brahe Fellow}

\altaffiltext{4}{Department of Astronomy and McDonald Observatory, The University of Texas, 1 University Station C1402, Austin, Texas 78712-0259, U.S.A.; wheel$@$astro.as.utexas.edu}

\altaffiltext{5}{Department of Physics, Texas A\&M University,
College Station, Texas 77843-4242, U.S.A.; wang$@$physics.tamu.edu}

\altaffiltext{6}{ESO - European Organisation for Astronomical Research
  in the Southern Hemisphere, Karl-Schwarzschild-Str.\ 2, 85748
  Garching b.\ M\"unchen, Germany; dbaade$@$eso.org; fpatat$@$eso.org}

\altaffiltext{7}{Departamento de Astronomía y Astrofísica, PUC
Casilla 306, Santiago 22, Chile; aclocchi$@$astro.puc.cl; jquinn$@$astro.puc.cl; pazelaya$@$astro.puc.cl}

\altaffiltext{8}{Department of Physics, Florida State University, Tallahassee, Florida 32306-4350, U.S.A.; pah$@$astro.physics.fsu.edu}

\begin{abstract}
  We report two spectropolarimetric observations of SN~2005hk, which
  is a close copy of the ``very peculiar" SN~2002cx, showing low peak
  luminosity, slow decline, high ionization near peak and an unusually
  low expansion velocity of only about 7,000 \kms. Further to the data
  presented by Chornock et al., (2006), at -4 days before maximum, we
  present data of this object taken on 9 November 2005 (near maximum)
  and 23 November (+ two weeks) that show the continuum and most of
  the spectral lines to be polarized at levels of about 0.2-0.3\%. At
  both epochs the data corresponds to the Spectropolarimetric Type
  D1. The general low level of line polarization suggests that the
  line forming regions for most species observed in the spectrum have
  a similar shape to that of the photosphere, which deviates from a
  spherical symmetry by $<10\%$. In comparison with spectropolarimetry
  of Type Ia and Core-collapse SNe at similar epochs, we find that the
  properties of SN~2005hk are most similar to those of Type Ia SNe. In
  particular, we find the low levels of continuum and line
  polarization to indicate that the explosion mechanism is
  approximately spherical, with homogeneous ejecta (unlike the
  chemically segregated ejecta of CCSNe).  We discuss the possibility
  that SN~2005hk was the result of the pure deflagration of a white
  dwarf and note the issues concerning this interpretation.
\end{abstract}
 
\keywords{supernovae -- spectropolarimetry}
\setcounter{footnote}{0}
\section{Introduction}
\label{sec:intro}

\begin{figure}
\includegraphics[width=8.5cm, angle=270]{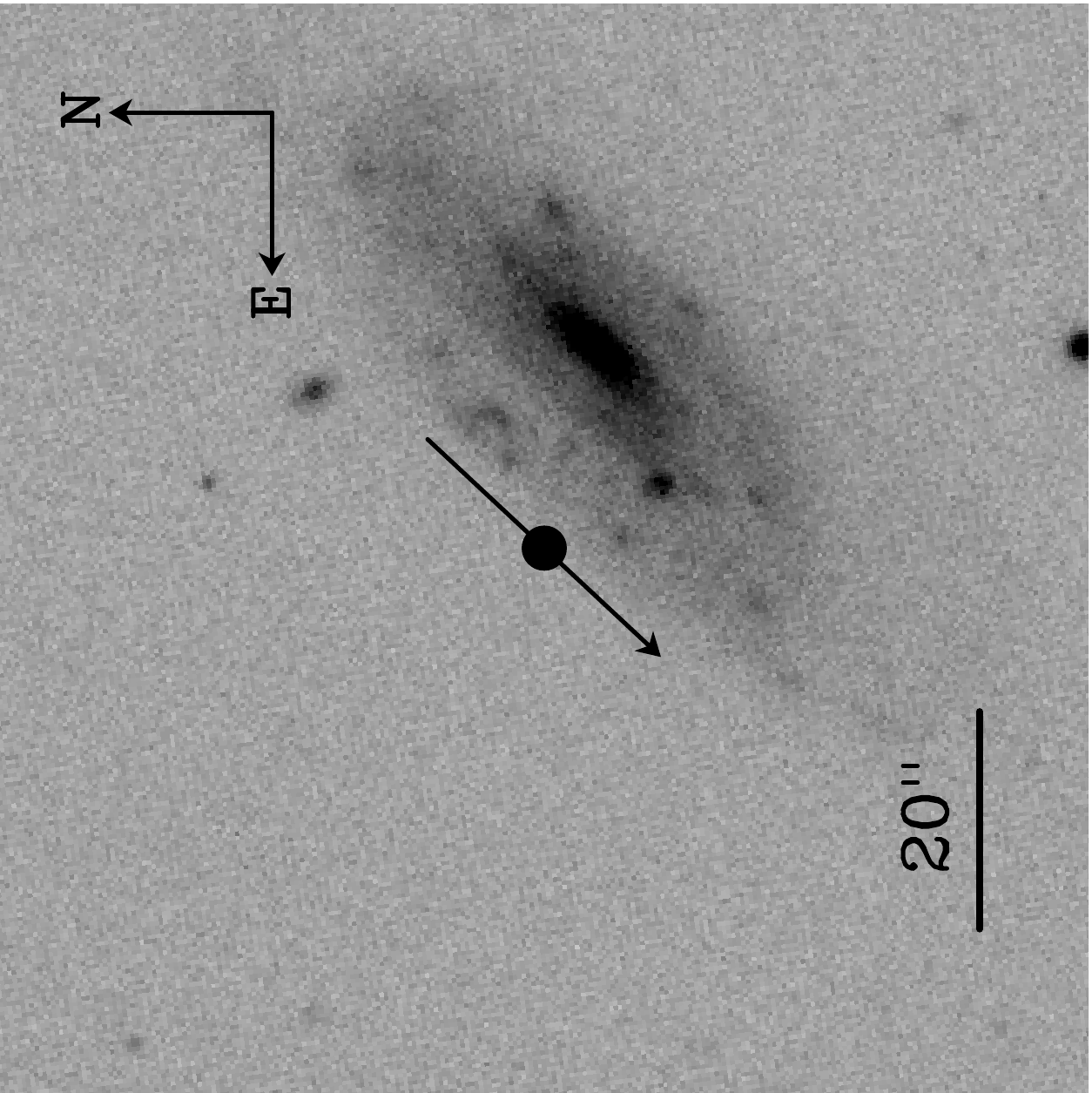}
\caption{SDSS $r'$-band image of the site of SN 2005hk, indicated by
  the filled circle, relative to its host galaxy UGC 272.  The orientation of the determined ISP component, at $\mathrm{PA=141\pm3\degr}$, is shown by the arrow.}
\label{fig:obs:hostimage}
\end{figure}

The nature of the complex explosion mechanisms thought to be behind
Supernovae (SNe), of all types, is expected to leave a signature in
the resulting geometry of the explosion.  The study of the polarized
light arising from SN explosions provides a unique and powerful method
for directly measuring asymmetries in SNe.  The power of
spectropolarimetry is demonstrated by its application to the study of
the shapes of SNe at large distances, providing an effective spatial
resolution far superior to any current or currently planned imaging
technique \citep[see][for a review]{2008ARA&A..46..433W}.\\
In SN photospheres, the predominant form of opacity is due to electron
scattering, which is also a polarizing process.  In a spherical
symmetry, light originating from different quadrants of the
photosphere have opposite but equal polarization components, leading
to a zero net observed polarization.  In the presence of asymmetries,
the cancellation of the polarization components arising from different
portions of the ejecta is incomplete, leading to an observable
polarization whose magnitude is related to the degree of the
asymmetry.  In addition to the polarization of continuum light, the
absorption due to line forming species exterior to the photosphere
“cast shadows” on the photosphere.  If the photosphere is not
completely covered, a polarization signature will be associated with
wavelength-dependent absorption features, with the degree of
polarization related to the amount and orientation of the coverage of
the photosphere by the line forming region.\\
Previous polarimetric studies of SNe have shown that Type Ia SNe and
Core-collapse Supernovae (CCSNe) have different polarization
properties.  Type Ia SNe are observed to have generally low
polarizations, indicative of a close-to-spherical symmetry.  The
degree of polarization of these events decreases with time, as deeper
layers of ejecta are revealed, implying the ionization/ density structure becomes more
spherically symmetric with depth.  In contrast, CCSNe have been
observed with a wide variety of polarizations, generally larger than
those observed for Type Ia SNe.  In addition, the degree of
polarization is observed to increase with time for CCSNe, suggesting
the ejecta become more asymmetric at increasing depth.\\
In the classical observing modes of photometry and spectroscopy, Type
Ia SNe at certain phases can be confused upon casual inspection with Type
Ic SNe at other phases due to the common display of \ion{Fe}{2}
lines. Special care must be given when trying to determine the type
and the physics of ``peculiar" Type Ia and ``peculiar" Type Ic SNe.
The application of spectropolarimetry can provide additional
constraints, both in the abstract sense of the Stokes parameters and
in the inferred geometries, that can help separate events belonging to
the  different distinct classifications.\\
One particular class of SNe, which have been previously classified as
Type Ia SNe (albeit ``the most peculiar Type Ia";
\citealt{2003PASP..115..453L}), are the so-called “2002cx-like” SNe.
It has recently been suggested that members of this class may be
representative of Type Ia SN events that result from a
pure-deflagration \citep[see][and references
therein]{2004PASP..116..903B}
or instead may be peculiar, low-luminosity Type Ic SNe \citep{2009AJ....138..376F,2009Natur.459..674V}.\\
Here we report two epochs of spectropolarimetry of the 2002cx-like
SN~2005hk, acquired at approximately maximum light and two weeks later
using the Very Large Telescope (VLT) of the European Southern
Observatory (ESO).  SN~2005hk was discovered in UGC 272 by
\citet{2005IAUC.8625....1Q}.  The location of SN~2005hk, relative to
its host galaxy, is shown as Fig. \ref{fig:obs:hostimage}.  Early
spectroscopy was presented by \citet{2006AJ....132..189J},
\citet{2006PASP..118..722C}, \citet{2007PASP..119..360P} and \citet{2008ApJ...680..580S}.  They
identify the early high ionization (e.g. \ion{Fe}{3}) species typical
of SN~2002cx-like SNe, as well as the bright Type Ia SN 1991T \citep[see
also][]{serduke05}, and the low velocities characteristic of SN~2002cx
($\sim$ 6000 \kms).  The spectra of \citeauthor{2006AJ....132..189J}
and \citeauthor{2006PASP..118..722C} showed intermediate-mass elements
with weaker features than regular SNe Ia, but stronger than SN~1991T.
\citeauthor{2006PASP..118..722C}\ suggested that Ni and Co may be
present in the early spectra and hence in the outermost layers of the
ejecta. The red portions of the spectra show appreciable features of
\ion{O}{1} and \ion{Mg}{2} between 7500 and 9000\AA, whereas SN~1991T, a
possible analog, showed a basically featureless spectrum in that phase
and wavelength range. Both SN~2002cx and SN~2005hk showed similar fine
scale structure suggesting that the structure was systematic and hence
not the product of random clumping in the ejecta of either
event. \citet{2007PASP..119..360P} present extensive {\it
  u'g'r'i'BVRIJHK} photometry and optical spectroscopy of SN~2005hk,
including the spectral data of \citeauthor{2006AJ....132..189J} and
\citeauthor{2006PASP..118..722C}.  Importantly, they note that the
light curves of these two events are very similar, with neither
showing the typical secondary maxima in R and I displayed by normal
SNe~Ia, but light curves that are also distinctly different than
under-luminous SNe~Ia such as SN~1991bg. In addition, the bolometric
light curve, including Swift UV data, shows a
rather slow light curve ($\Delta m_{15}(B) = 1.56$; \citealt{2007PASP..119..360P}), uncharacteristic of under-luminous SN~Ia and similar to normal Type Ia SNe over a magnitude brighter at maximum.\\
\citet{2006PASP..118..722C} presented spectropolarimetry of SN~2005hk,
at 4 days prior to our first observation. They determined that the
continuum was polarized $\sim 0.4\%$ in the red, with the data
represented by a dominant axis. They also identified a weak modulation
of the polarization of the \ion{Fe}{3}
line at $\mathrm{\lambda5129}$.\\
In Section \ref{sec:obs} we present our spectropolarimetric
observations and our adopted reduction procedure.  The results from
these observations are presented in \S\ref{sec:results} and these
results are discussed in
\S\ref{sec:disc}.  In \S\ref{sec:conc} we present our conclusions.\\

\section{Observations}
\label{sec:obs}
Spectropolarimetry of SN~2005hk was acquired using the ESO-VLT on 2005
Nov 9 and 23, with the FORS1 instrument in polarization PMOS mode.  A
log of these observations is presented in Table \ref{tab:obs:log}. The
data were reduced following the scheme outlined in \citet{maund05bf}.
At each epoch, the observations consisted of four exposures with the
half-wavelength retarder plate positioned at four position angles
$0\degr$, $45\degr$, $22.5\degr$, and $67.5\degr$, to derive the two
linear Stokes parameters $Q$ and $U$\footnote{We adopt the notation of
    \citet{2008ARA&A..46..433W}, where the normalized Stokes
    parameters are given by $Q=\widehat{Q}/I$ and $U=\widehat{U}/I$.}.  Both observations
were conducted with the 300V grism, without an order separation
filter, providing a spectral resolution of $12.5\AA$ as measured from
arc line calibration data.  Telluric lines were removed from flux
spectra in the standard manner.  Flux standard stars were observed,
with the full polarimetry optics in place, to calibrate the flux
levels of the SN data and remove telluric features.  The telluric
features are not expected to affect the polarimetry, except to
decrease the level of signal-to-noise across the absorption, provided
there is no significant change in the strength of absorptions between
exposures with the retarder plate rotated by $45\degr$.\\
The instrumental stability between the observations at the two epochs
was assessed by calculating the difference between the instrumental
signature corrections $\epsilon_{Q}$ and $\epsilon_{U}$ \citep{2008A&A...481..913M}.  At both epochs the
difference was found to be $0\pm0.1\%$, implying a limit on the
instrumental stability and the measurement of the Stokes parameters to
$0.1\%$.  This is consistent with previous reported measurements of
the instrumental signature corrections and limits on stability of the
VLT FORS instrument \citep{2009ApJ...705.1139M}.\\
The degree of polarization was corrected for biases due to
observational noise using the equations derived in
\citet{1997ApJ...476L..27W}, and, additionally, using a Monte Carlo
simulation of the FORS1 instruments in the manner of
\citet{2006PASP..118..146P} and \citet{2008A&A...481..913M}.  The
wavelength scales were corrected for the heliocentric recessional
velocity of the host galaxy of 3895
$\mathrm{km\,s^{-1}}$.\footnote{http://nedwww.ipac.caltech.edu/}\\

\section{Observational Results}
\label{sec:results}

\subsection{The Distinguishing Spectral Features}
\label{sec:results:spectrum}

\begin{figure}
\includegraphics[width=7cm, angle=270]{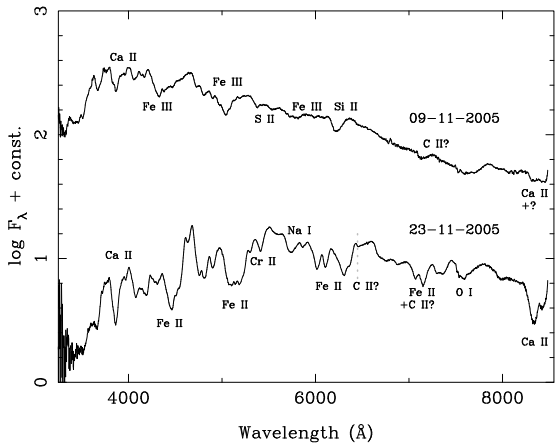}
\caption{Flux spectra of SN~2005hk at 09 Nov 2005 and 23 Nov 2005, or
  approximately at B light curve maximum and 14 days after maximum.
  Line identifications are based on those provided by
  \citet{2006PASP..118..722C}, \citet{2007PASP..119..360P},
  \citet{2004PASP..116..903B} and fits to the spectra using the SYNOW
  code.  The wavelength scale has been corrected for the recessional
  velocity of the host galaxy.}
\label{fig:obs:spec}
\end{figure}

The spectroscopic properties of SN~2005hk have already been described
in some detail in other works
\citep{2006PASP..118..722C,2007PASP..119..360P}.  Here the prominent
spectroscopic features, which characterise SN~2005hk at the epochs at
which spectropolarimetry was acquired, will be briefly reviewed to
orient  the reader.  The flux spectra from our two spectropolarimetric observations are presented as Fig. \ref{fig:obs:spec}.\\
Our data from 2005 November 9 are interleaved with the coverage presented by
\citet{2007PASP..119..360P} who give spectra from 7 and 14 November
that they assign to -3 and +4 days with respect to B maximum which
they determine to be JD2,453,685.1. This means our data are from very
near B maximum, perhaps a day earlier. We confirm that the spectral
features are very similar to those of SN~2002cx at the same epoch
\citep{2003PASP..115..453L}. Furthermore, \citet{2006PASP..118..722C}
acquired a flux spectrum of SN~2005hk in the course of acquiring
spectropolarimetry at November 5.4, 4 days prior to our first
observation.  \citeauthor{2006PASP..118..722C} performed line
identifications using the SYNOW\footnote{http://www.nhn.ou.edu/~parrent/synow.html} code for their spectrum. They found
that the strongest lines were from \ion{Fe}{3} with other lines of
intermediate mass elements being present, but relatively weaker. They
obtained a better fit in the blue by adding lines of \ion{Ni}{2},
\ion{Co}{2}, \ion{Ti}{2} and \ion{C}{3}. They used an excitation
temperature of 10,000 K for the low-ionization stage elements and
15,000 K for the higher ionization stages, \ion{Fe}{3} and
\ion{Si}{3}.  Our flux spectrum from the 2005 November 09 observation,
despite being acquired 4 days later, requires similar identifications
to those made by \citeauthor{2006PASP..118..722C}.  In the red portion
of the spectrum, we identify the \ion{Ca}{2} IR triplet, although the
synthetic spectra of \citeauthor{2006PASP..118..722C} suggested that
at the epoch of their observations this was mostly due to \ion{O}{1}
and \ion{Mg}{2}.  The feature at $\sim 7600\AA$ is identified by
\citet{2007PASP..119..360P} as \ion{Fe}{2}, while
\citet{2006PASP..118..722C} suggested it was a mixture of \ion{O}{1}
and \ion{Mg}{2}.  \citet{2003PASP..115..453L} observed the appearance
of \ion{O}{1} $\lambda 7774$ after maximum in spectra of SN~2002cx,
which suggests that at 0 days for SN~2005hk this feature could be
at least partially due to \ion{O}{1}.\\
Our spectrum from 2005 November 23 corresponds to an epoch two weeks
after B maximum. Our data are very similar to spectra of that date
presented by \citet{2007PASP..119..360P}, but our data have higher 
signal to noise and broader wavelength coverage. The spectrum of \citealt{2007PASP..119..360P} from November 25
(+15 days) has similar wavelength coverage to ours of November 23 and
remains very similar to our flux spectrum.  At this epoch, the
underlying continuum is cooler with a relative paucity of flux in the
blue compared to the spectrum of 9 November. The spectrum is
characterized by the growth in amplitude of small scale features.
SYNOW was used by us, in conjunction with previously published line
identifications, to determine line identifications specific to the
spectrum at this epoch.  The degree of blending of multiple lines is
lower at the second epoch, due to the relative decrease in the
expansion velocity.  The truncation of the red-side of the
\ion{O}{1} $\lambda 7774$ emission line profile at $\mathrm{7980\,\AA}$ may be evidence for \ion{Mg}{2},
but it is located close to strong telluric absorption features.  The
identification of \ion{Si}{2} $\lambda 6355$ at +14 days is debatable,
and is likely to be due to an increase in \ion{Fe}{2} $\lambda\lambda
6456,6518$ \citep{2003PASP..115..453L}.  \citet{2004PASP..116..903B}
similarly found that \ion{Si}{2} was no longer required to fit spectra
of SN~2002cx at a similar epoch.  The photospheric velocity was
estimated, using lines of \ion{Fe}{2}(42), to be $\sim -7000$\kms.\\
At the second epoch, there are also two distinct absorption features
that are likely to be caused by the \ion{Ca}{2} IR triplet.  The
\ion{Ca}{2} IR triplet was identified at 12 days after maximum in the
spectrum of SN~2002cx \citep{2004PASP..116..903B}. The two features
might be due to partially resolving the triplet in this low velocity
photosphere ($-5650$ \kms), or there could be two kinematically separate components
such as those observed in normal Type Ia SNe
\citep{2005ApJ...623L..37M}.  If the latter, the absorption minima
correspond to velocities of $-8400$ and $-5500$ \kms, respectively, taking
the mean wavelength of the \ion{Ca}{2} blend to be 8579\AA. The bluer
of these two components is moving somewhat more rapidly than the
typical range identified for photospheric features for Type Ia SNe at
similar epochs \citep{2005ApJ...623L..37M}.  Comparison with SN~2002cx
at 12 days \citep{2004PASP..116..903B} leads to the identification of
the strong absorption observed at 5730\AA\ as \ion{Na}{1} intrinsic
to the SN, corresponding to a velocity of $-8150$\kms (which is
similar to the high velocity \ion{Ca}{2} feature).\\
At the first epoch, there is a feature at $\sim 7200\AA$.  \citet{2006PASP..118..722C} argue it is unlikely to be due to
\ion{C}{2} due to absence of other stronger lines that should occur in
the wavelength range of these observations (e.g. $\lambda 6578$).
Rather \citet{2008ApJ...680..580S} suggests the $7200\AA$ may arise
from \ion{Fe}{2} $\lambda 7308$. \citet{2004PASP..116..903B} could not identify \ion{C}{1} features in
spectra of SN~2002cx.  At the second epoch, however, there is a suggestion of
an absorption (blended with an \ion{Fe}{2} line giving rise to a flat
topped emission line profile for that line) which is consistent with
the expected position of \ion{C}{2} $\lambda 6578$ with a velocity of
$-5500$ \kms\ (similar to the low velocity component of \ion{Ca}{2}). 

\subsection{The Polarization of SN~2005hk}
\label{sec:res:pol}

\begin{figure}
\includegraphics[width=9cm, angle=270]{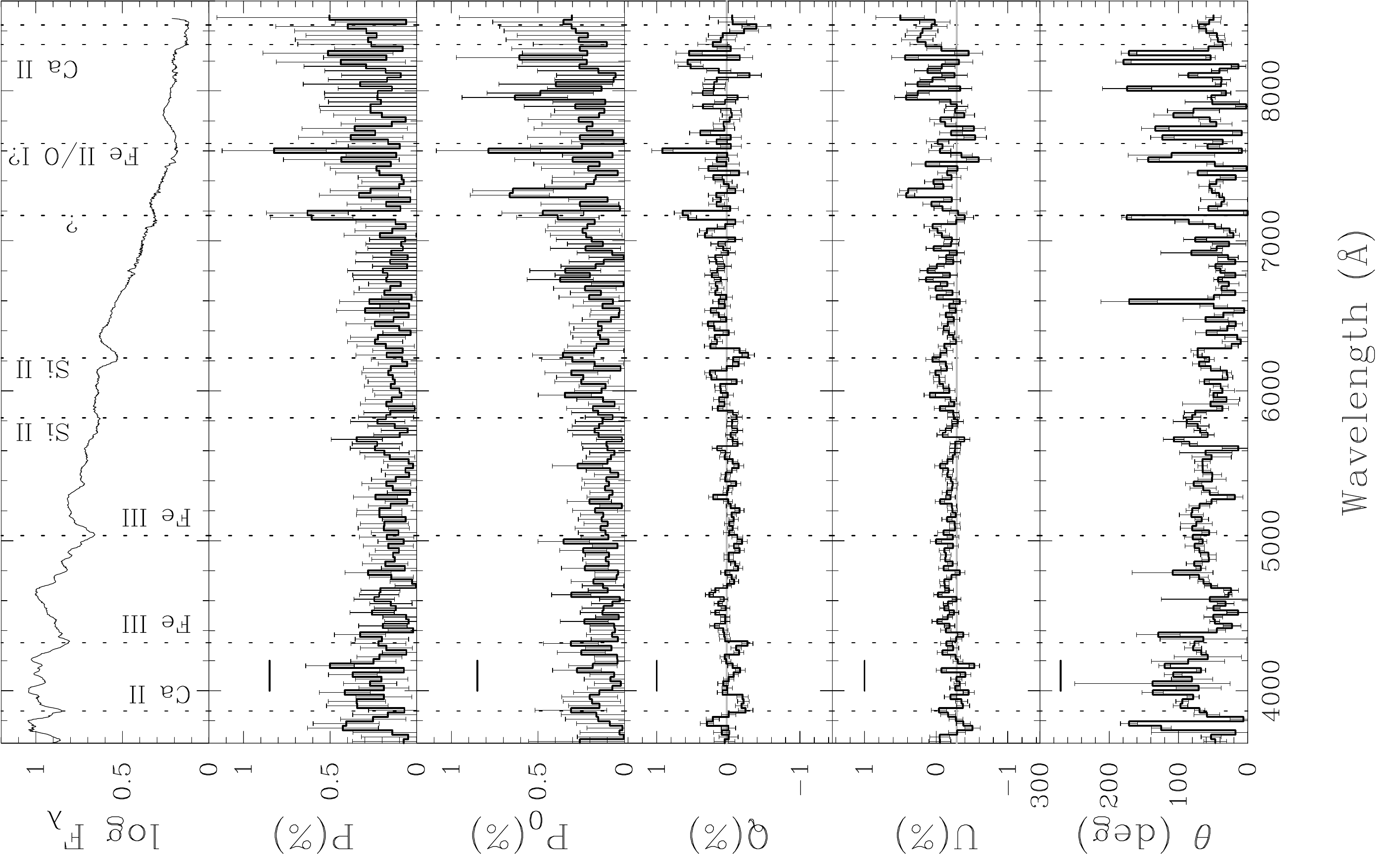}
\caption{Spectropolarimetry of SN~2005hk on 2005 Nov. 9.  The six
  panels (from top to bottom) give: the observed flux spectrum, in
  logarithmic flux units ($\mathrm{ergs\,s^{-1}\,cm^{-2}\,\AA^{-1}}$);
  the observed polarization spectrum ($P$); the observed polarization
  spectrum corrected for the ISP component ($P_{0}$); the Stokes $Q$
  and $U$ spectra uncorrected for the ISP (the value of the ISP is
  shown for each parameter by the grey line); and the polarization
  angle $\theta$.  Line identifications are provided in the top panel,
  and are based on the identifications by \citet{2006PASP..118..722C},
  \citet{2004PASP..116..903B}, \citet{2007PASP..119..360P} and
  comparison with synthetic spectra produced using the SYNOW code.
  The data have been rebinned to 30$\mathrm{\AA}$\ for clarity.  The
  wavelength range selected for the determination of the ISP are shown
  by the heavy lines.}
\label{fig:obs:panela}
\end{figure}

\begin{figure}
\includegraphics[width=9cm, angle=270]{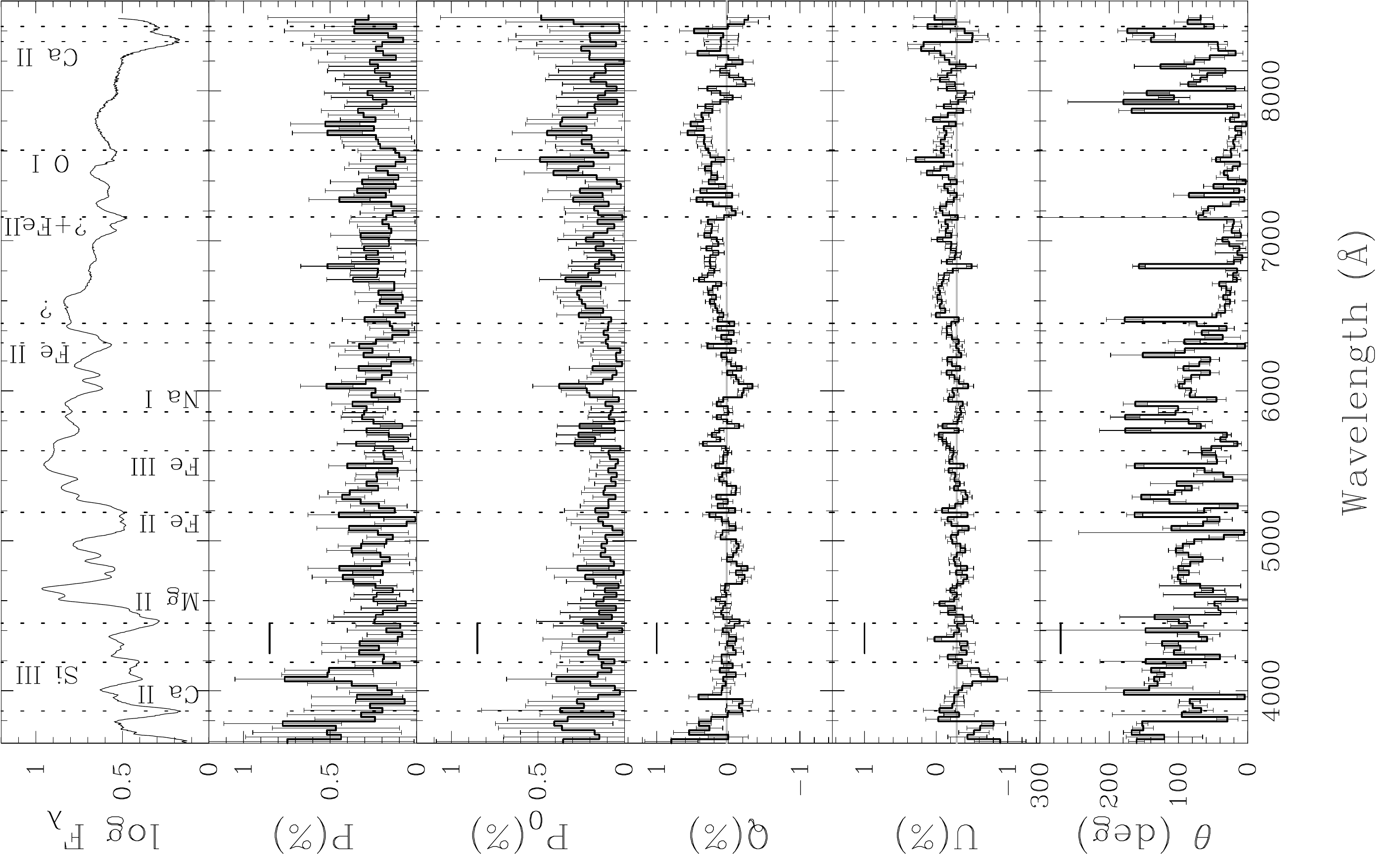}
\caption{Same as Fig. \ref{fig:obs:panela}, but for spectropolarimetry of SN~2005hk on 2005 Nov. 23.}
\label{fig:obs:panelb}
\end{figure}

\begin{figure}
\includegraphics[width=8.5cm, angle=270]{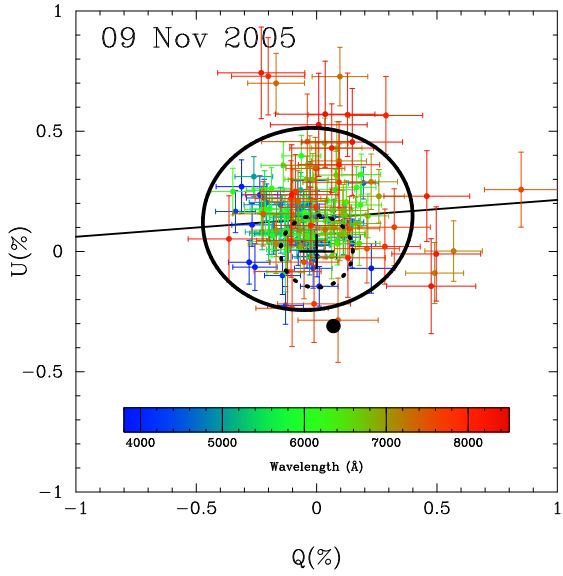}
\caption{Spectropolarimetry of SN~2005hk on 2005 Nov. 9, on the Stokes
  $Q-U$ plane.  The data have been binned to 30$\mathrm{\AA}$ for
  clarity, and have been corrected for the ISP component, which is
  indicated on the Stokes plane by the filled black circle, with the
  radius of the circle corresponding the measured uncertainties.  The
  data are colour coded according to wavelength following the scheme
  of the colour bar.  The origin of the Stokes plane is indicated by
  the cross ($+$).  The black dashed circle represents the null polarization
  envelope, at 6000$\mathrm{\AA}$.  The heavy black line indicates the
  best fit to the
  ``dominant axis.''  The ellipse shows the apparent axial ratio
  and orientation of the data, as determined using the Principal
  Components Analysis, with the size of the ellipse arbitrarily
  scaled.}
\label{fig:obs:qupanela}
\end{figure}

\begin{figure}
\includegraphics[width=8.5cm, angle=270]{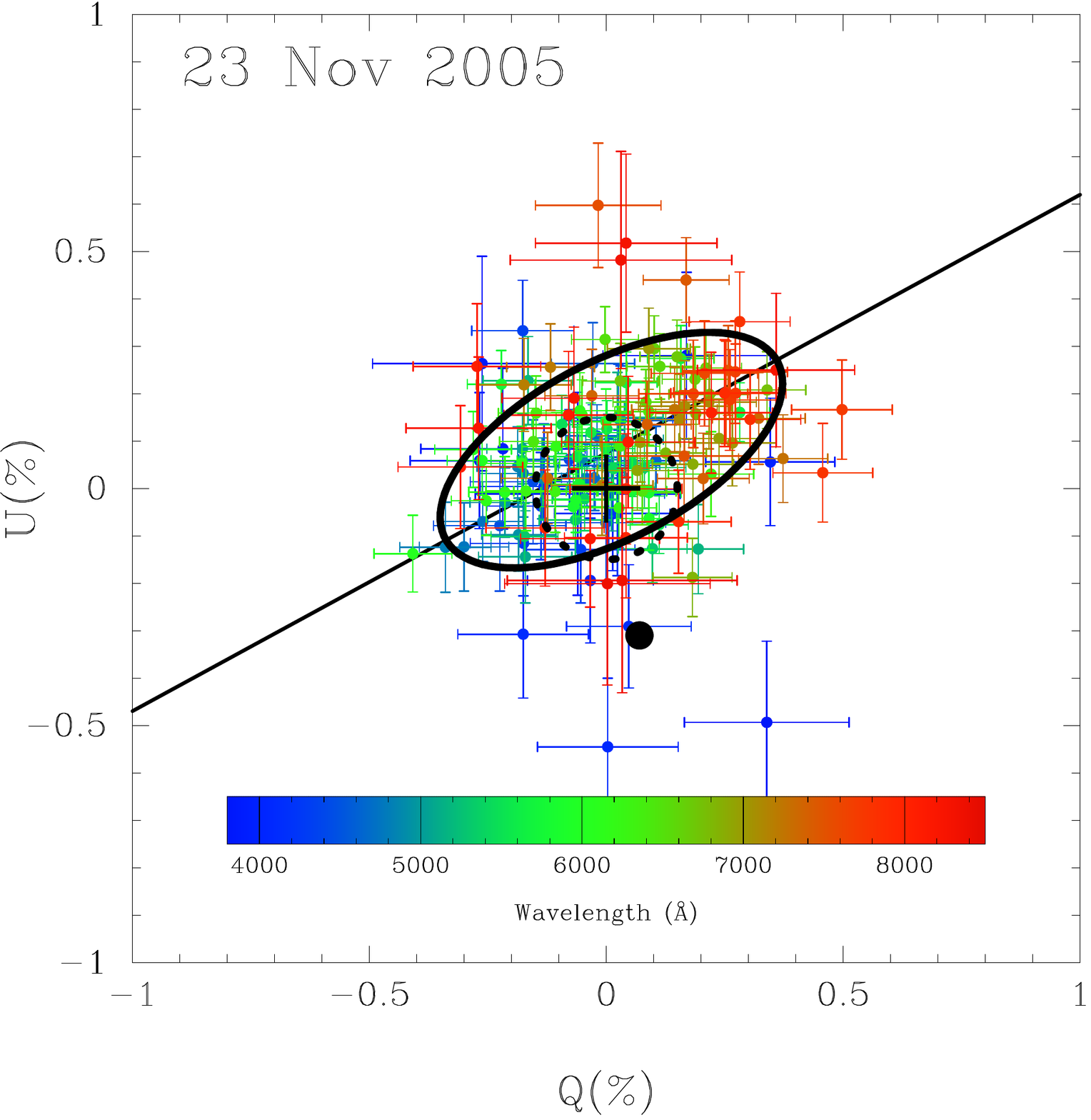}
\caption{Same as Fig. \ref{fig:obs:qupanela} but for spectropolarimetry of SN~2005hk at 2005 Nov. 23.}
\label{fig:obs:qupanelb}
\end{figure}

Spectropolarimetry of SN~2005hk at the two epochs is presented in Figs. \ref{fig:obs:panela} and \ref{fig:obs:panelb}.\\
The correct measurement of the intrinsic polarization of SNe requires
the subtraction of the polarization contribution of the interstellar
medium, both in the Galaxy and in the host.  In the first instance, a
general limit can be placed on the degree of the Interstellar
Polarization (ISP) by the amount of reddening measured towards SN
2005hk.  For Galactic-type dust the maximum degree of the ISP is
related to the reddening by the limit $p_{ISP} < 9\times E(B-V)$
\citep{1975ApJ...196..261S}.  The degree of Galactic foreground
reddening in the direction of SN 2005hk has been reported to be
$E(B-V)=0.022$ \citep{schleg98}. The limit on ISP arising in the foreground is $0.2\%$.\\
The reddening towards SN~2005hk can be estimated, if we assume that the narrow \ion{Na}{1} D absorption component
arises in the interstellar medium (ISM) of the host. The equivalent width of the \ion{Na}{1} D absorption
feature at the first epoch was measured to be
$W_{\lambda}=0.30\pm0.02$.  This value is smaller than the value
reported by \citet{2006PASP..118..722C}, for observations 4 days prior
to our first observations, although the difference between the two
measured values is $\sim 2\sigma$. We note
\citeauthor{2006PASP..118..722C} do not quote errors on their
determined equivalent width.  It is not clear if this difference is
within the measurement errors or if it reflects evolution triggered by the SN \citep{2007Sci...317..924P}.  We
follow \citet{2006PASP..118..722C}, who adopt the results of
\citet{1997A&A...318..269M} for evaluating the relationship between
\ion{Na}{1}D absorption and the reddening.  This relationship is based
on Galactic-type extinction laws \citep[e.g.][]{ccm89}.  It is,
therefore, implicitly assumed that the reddening component, measured
from the narrow \ion{Na}{1} absorption component in the host galaxy,
follows a similar law.  As the \ion{Na}{1} D component lines are not
resolved in our data, we also adopt a range of flux ratios between
\ion{Na}{1} D1 and D2 of 1.1 to 2 \citep{2006PASP..118..722C}.  This
limits the reddening of the host to $0.05<E(B-V)<0.07$.  The total
reddening, for Galactic type reddening and ISP law, yields a
constraint on the total
ISP of $p_{ISP}<0.8\%$.\\
\citet{2001ApJ...556..302H} present a technique for the determination
of the ISP for specific wavelength ranges which have null intrinsic
polarization.  This assumes that, for
these specific wavelength ranges, there is sufficient overlap between
groups of lines such that there is complete intrinsic depolarization
over these wavelengths.  \citeauthor{2001ApJ...556..302H} suggest that
the wavelength region of 4800-5600\AA\ is intrinsically
depolarized, where line blanketing opacity dominates over electron
scattering opacity.  Any polarization, observed over this wavelength
range, is expected to be due to the ISP alone.  For SN 2005hk, at both
epochs, the lack of significant overlap and blending of lines leads to
the complex behaviour of the Stokes parameters.
\citet{2006PASP..118..722C} find that the wavelength range of
4000-4200\AA\ satisfies the criteria for intrinsic depolarization,
namely significant line blanketing.  The observation of
\citeauthor{2006PASP..118..722C} at November 5.4 is sufficiently close
to our observation at the first epoch that we observed similar
features in the polarization spectrum and the Stokes parameter
spectra.  Importantly, we too determine that this wavelength range
provides the best approximation to a line blanketed regime.  The
measured Stokes parameters across this range at the first epoch are
${Q}=0.04\%\pm0.03$ and ${U}=-0.32\%\pm0.03$.  The
uncertainties on these quantities are the combined systematic and
statistical uncertainties.\\
The spectral evolution of SN 2005hk by the second epoch, with the
increase in the strengths of individual \ion{Mg}{2} and \ion{Fe}{3} lines in the
range of 4000-4200\AA, leads to the presence of intrinsic polarization
(${Q}=0.12\%\pm0.06$ and
${U}=-0.47\%\pm0.060$). Instead, we identify the region of
4250-4450\AA\ as being spectroscopically similar to a line blanketed
regime with no unblended line structure.  Over this wavelength range
the average values of the Stokes parameters are measured as
${Q}=0.09\%\pm0.03$ and ${U}=-0.30\% \pm 0.03$.  The
measured Stokes parameters, over the depolarized wavelength range,
agree with those determined at the first epoch.\\
We adopt values of the ISP Stokes parameters of
$Q_{ISP}=0.07\%\pm0.03$ and $U_{ISP}=-0.31\% \pm 0.03$, for both the
Galactic and host ISP components, which are
approximately consistent with the values measured by
\citeauthor{2006PASP..118..722C} (who calculate separate Galactic and host components). The value of the ISP relative to
the observed Stokes parameters is shown on Figs. \ref{fig:obs:panela} and \ref{fig:obs:panelb}. These values correspond to a
polarization of $p_{ISP}=0.32\%\pm0.03$ with an angle of
$\theta_{ISP}=141\degr\pm3$.  The orientation of the ISP component is
shown on Fig. \ref{fig:obs:hostimage}.  As previously observed for
other SNe \citep{my2001ig,2009A&A...508..229P,2010A&A...510A.108P}, the ISP component is aligned
with the spiral arm pattern at the SN location, due to alignment of
magnetic field lines and, hence, dust grains along
spiral arms \citep{1987MNRAS.224..299S}. \\
Given the limited wavelength range over which it can be determined, it
is not clear if and how the ISP component varies with wavelength.
Although a standard Serkowski-type law could be applied, it is unclear
what the appropriate values of $\lambda_{max}$ and $p_{max}$ would be.
Furthermore, it has been observed for a number of SNe with significant
ISPs that the ISP does not necessarily even follow a Galactic-Serkowski
type law of wavelength dependence \citep[e.g. SN2006X;][]{2009A&A...508..229P}.  Given
the small value of the ISP, we subtract, therefore, a constant ISP
component from the observed data, to recover the intrinsic
polarization of the SN (as shown by $P_{0}$ in
Figs. \ref{fig:obs:panela} and
\ref{fig:obs:panelb}), under the assumption that any wavelength
dependence of the ISP is smaller than the uncertainties derived above.\\
On the Stokes $Q-U$ plane (see Figs. \ref{fig:obs:qupanela} and
\ref{fig:obs:qupanelb}), the ISP component is located close to the centroid of the SN~2005hk data.  The calculated weighted centroids for the ISP-corrected data $\left( \overline{Q} , \overline{U} \right)$ are listed in Table \ref{tab:obs:domaxis}.\\
After correction for the ISP, the intrinsic polarization of SN~2005hk
is observed to be extremely low across the observed wavelength range.
The degree of polarization of a number of features observed to be
relatively strongly polarized in the uncorrected polarization spectrum,
are no longer as polarized once the data have been corrected for the
ISP.  Utilising the data across the range $6600-7000\AA$ as
representative of a region of pure continuum emission, the level of
the continuum polarization is measured as $0.17\pm0.05\%$ and
$0.19\pm0.05\%$ at the first and second epochs, respectively. At the
second epoch the assumption that this wavelength range represents a
true continuum region is most likely invalid, due to the presence of a
number of minor, broad undulations which are most likely due to the
lines (rather than incorrect flatfielding of the flux spectrum).  This
wavelength range avoids any polarization associated with the feature
at $7200\AA$, and is less than the continuum polarization measured by
\citet{2006PASP..118..722C} at 2005 Nov 5.4 of $0.36\%$. This level of
polarization is consistent with a spheroidal configuration
with axial ratio $>0.9$ \citep{1991A&A...246..481H}.\\
The significance of the polarization signal is given by $(S/N)_{P}=
P/\sigma_{P}= \sqrt{N/2}P(S/N)$ \citep{2006PASP..118..146P}.  At both
epochs there are no features, at either $15\AA$\ or $30\AA$\ binning,
which are detected in the intrinsic polarization spectrum $P_{0}$ at
$>3\sigma$ and only a small number of bins detected at $>2\sigma$.
\citet{2006PASP..118..722C} argue that in their data from 5 November,
only the line of \ion{Fe}{3} $\lambda 5129$ is polarized and at a
rather small level. On inspection, their data suggest that \ion{Fe}{3}
$\lambda 4404$ is also somewhat polarized.   There is another feature
of similar amplitude showing a minimum at $\lambda 4800$ in the
polarized flux that has no strong component in the total flux
spectrum, but is likely to correspond to \ion{Si}{2} $\lambda 5051$.\\
 Our data at the first
epoch also show possible modulation in the polarization spectrum
associated with \ion{Fe}{3} $\lambda 5129$, as well as \ion{Si}{2} $\lambda 6355$ and
the $7200\AA$\ feature.  At redder wavelengths there are a number of
peaks which are associated with telluric features, suggesting that
these were variable over the course of the observations.  The apparent
significance of these features is therefore overestimated, as the
possible variability of telluric features was not considered in the
reduction process and not accounted for in the final error analysis.\\
By the second epoch, there are possible peaks in the intrinsic polarization associated
with the absorptions of the \ion{Fe}{2} multiplet 42, the intrinsic
\ion{Na}{1} feature and the absorption of \ion{Fe}{2} $\lambda 6147$.\\
The absence of a significant signal in the polarization spectrum does
not necessarily imply that the intrinsic polarization of spectral
features is also not significant.  \citet{1997ApJ...476L..27W} and
\citet{2002PASP..114.1333L} studied the change in polarization $\Delta
P$ across particular spectral features, by comparing the modulation in
the Stokes $Q$ and $U$ parameters from the approximate ``continuum''
level and at the wavelength of absorption minimum, where:
\begin{equation}
\Delta P = \sqrt{\left(Q\left(\lambda_{cont}\right)-Q\left(\lambda_{line}\right)\right)^{2}+\left(U\left(\lambda_{cont}\right)-U\left(\lambda_{line}\right)\right)^{2}}
\end{equation}
$\Delta P$ is independent of the choice of the ISP component.  The values of $\Delta P$ for particular spectral features, at both epochs, are presented in Table \ref{tab:res:delp}.\\
Even when assessed by modulation of the Stokes parameters, the data at
both epochs show few significant features.  At the first epoch,
\ion{Ca}{2} H\&K, \ion{Fe}{2} and, importantly, \ion{Si}{2} $\lambda
6355$ show significant modulation, whereas at the second epoch the
significant modulation is associated with the, now stronger,
\ion{Fe}{2} lines.  The interpretation of the data presented in Table
\ref{tab:res:delp} requires some caution, however, as the selection of
$\lambda_{cont}$ and $\lambda_{line}$, by eye, can lead to a bias in
favour of selecting those wavelengths that yield the maximum change in
the Stokes parameters.
The principal differences between the data presented by \citeauthor{2006PASP..118..722C} at -4 days, compared to our data at 0 days, is due to much larger S/N achieved in their observation.  We estimate that at $7200\AA$, at 50\AA\ binning, \citeauthor{2006PASP..118..722C} achieved a final $S/N \approx 1400$, whereas our observation reached $S/N\approx 390$.\\
At the first epoch, two sets of two contiguous bins (at 30\AA) are
observed at both the 7200\AA\ feature and immediately redward at
7315\AA.  An obvious line feature is associated with the
former spectral line in the flux spectrum.   The latter feature
corresponds to a telluric feature that, although it should be equally
cancelled between the two beams at each retarder plate position, may
appear polarized due to variability in
the telluric line between the pair of observations used to derive each
Stokes parameter. \\
On the Stokes $Q-U$ plane (see Figs. \ref{fig:obs:qupanela} and
\ref{fig:obs:qupanelb}), the data at both epochs assume an approximate
elliptical distribution, with no major structure deviating from the
main concentration of points.  At the first epoch, the data have no
obvious preferred direction.  By the second epoch the data are more
extended across the $Q-U$ plane, although again with no major
deviations from the main concentration of data points.  From
Fig. \ref{fig:obs:qupanelb} it is also evident that, by the second
epoch, the degree of polarization has become more wavelength
dependent, with data at bluer wavelengths being concentrated at lower values
of $Q$ and $U$ than the data at the red extreme of the observed
wavelength range. Our lower $S/N$ data does not show the wavelength
dependent elongation to the same degree as observed at -4 days, for
which a dominant axis is clear to the eye. The behaviour of the data
on the Stokes $Q-U$ plane, at both epochs, is quantified and discussed
in Section \ref{sec:res:decomp}.

\subsection{Decomposition of the Stokes parameters}
\label{sec:res:decomp}

The determination of the dominant axis, and the decomposition of the
Stokes vector, permits a correction for the random position angle at
which target SNe have been observed, as well as determination of the
principal axis of symmetry of the data, the relative orientations of
the different line forming species and, ultimately, the underlying
geometry.  The dominant axis forms a line across the Stokes $Q-U$
plane, and the decomposition process involves rotating the data so
that the principal rotated Stokes axis is aligned with the axis of
greatest variance for the data on the Stokes plane.  The polarization signal is measured in the two orthogonal directions, which are
effectively the  ``rotated Stokes parameters'', along the dominant axis
($P_{d}$) and along the orthogonal axis ($P_{o}$).  A polarization
component in the orthogonal direction permits the identification of
deviations from the single axial
symmetry represented by the dominant axis.  \\
Here we explore two complementary techniques for determining the
dominant axis: least-squares fitting of a straight line and a
principal components analysis of the data on the Stokes $Q-U$ plane.
Importantly, these two different analyses {\it should} give identical
measures of the dominant axis, but provide different parametrisation
of the data on the Stokes $Q-U$ plane.  This latter point is
particularly crucial for applying the SN spectropolarimetric
classification scheme of \citet{2008ARA&A..46..433W}.\\

\subsubsection{Least-squares fit to the data}
\label{sec:res:lsf}
Straight lines, of the form $U=\alpha+\beta Q$, were fit to the data
of both epochs, using a weighted least-squares fitting technique,
where the data were weighted according to the errors in both $Q$ and
$U$ \citep{1992nrca.book.....P}.  In order to confirm the validity of
this technique, the form of the straight line was reversed and the best fit was calculated
for $Q = 1/\beta U -\alpha/\beta$, giving identical results.  The value of $\beta$ yields the
angle of the dominant axis with respect to the Stokes $Q$-axis,
through which the data are to be rotated to decompose the Stokes
vector onto the dominant and orthogonal axes.  For a single axial
symmetry, the dominant axis will pass through the origin of the Stokes
$Q-U$ plane.  In instances where the data are offset from the origin
(e.g. by a wavelength-independent continuum polarization component)
the offset will be {\it partially} manifested in the
measured $\alpha$ parameter.\\
Dominant axes were calculated for the data at both epochs over two
different wavelength ranges: 3700 to 8600\AA\ and 4000-7000\AA.  The
former range wavelength encompasses the entire wavelength range of the
observed data, whereas the latter excludes regimes with poor levels of
signal-to-noise ($S/N \leq 200$).  The regions of low signal-to-noise
arise at the blue and red extremes of the data, due to line-blanketing
and the blue spectral energy distribution significantly reducing the
observed flux in the blue and red,
respectively, and the declining detector response (in the red).\\
The results of the least-squares analysis and the properties of the
determined dominant axes are presented in Table \ref{tab:obs:domaxis}.
The values of $\alpha$, $\beta$, and the polarization angle of the
dominant axis ($\theta_{dom}=\frac{1}{2}\arctan \beta$) were found to
be relatively insensitive, within the uncertainties, to the choice of
the wavelength range over which the dominant axis was calculated.  The
dominant axes are shown on the Stokes $Q-U$ plane on
Figs. \ref{fig:obs:qupanela}
and \ref{fig:obs:qupanelb}.   The large values of $\chi^{2}$ for the
straight line fit, at both epochs, indicate that the data are poorly
described by a straight line, with a significant polarization
component in the orthogonal direction.  At the first epoch, simple
inspection of the data on the Stokes $Q-U$ plane would suggest that
the data are evenly distributed in angle about the centroid (as is
illustrated in Section \ref{sec:res:pca}).  At the second epoch, a
dominant axis is more clearly observed.\\

\subsubsection{Principal Components Analysis}
\label{sec:res:pca}

In order to better understand and quantify the distribution of the
data on the $Q-U$ Stokes plane and, in particular, the degree of the orthogonal polarization component, a principal component analysis of the
data was conducted.\\
The elements of the $2 \times 2$ $weighted$ covariance matrix take the form:
\begin{equation}
cov_{w}(X_{j},X_{k})=
\frac{
\sum_{i}\frac{1}{\Delta X_{ij}}\frac{1}{\Delta X_{ik}}\left(X_{ij}-\overline{X}_{j}\right)\left(X_{ik} -\overline{X}_{k}\right)}{\sum_{i}\frac{1}{\Delta X_{ij}}\frac{1}{\Delta X_{ik}}}\frac{N}{N-1}
\end{equation}
where $j,k=0,1$, $X_{0}=Q$, $X_{1}=U$
\citep[see][]{2003drea.book.....B}, $i=1...N$ and the weighted average of the
Stokes parameters are $\overline{X}_{j}=\sum_{i}\frac{X_{ij}}{\Delta
  X_{ij}^{2}}/\sum_{i}\frac{1}{\Delta X_{ij}^{2}}$, as shown in Table
\ref{tab:obs:domaxis}, over the $N$ wavelength bins.  The eigenvectors
and eigenvalues of the covariance matrix provide an alternative measure
of the angles of the dominant and orthogonal axes ($\theta_{e}/2$) and
the ratio of the
degrees of polarization in the dominant and orthogonal directions (an axial ratio for the resulting ellipse $b/a$), respectively.  This latter quantity can be used to describe an ellipse on the Stokes $Q-U$ plane which contains the data.  The results of the principal component analysis are presented in Table \ref{tab:obs:domaxis}, and corresponding ellipses are presented on Figs. \ref{fig:obs:qupanela} and \ref{fig:obs:qupanelb}.  As expected, the polarization angle of the dominant axis determined using this analysis agrees, within the uncertainties, with the result from the least-squares fit.\\
The non-zero values of $b/a$ determined from the principal components
analysis show that the data at the two epochs are inconsistent with both 
straight lines and null polarization.  This suggests that, although the polarization
measured at both epochs is low, we have measured a real polarization
signal for SN~2005hk.  The ratio of the dominant and orthogonal
components measured at the second epoch is smaller than
that measured at the first, implying that at the first epoch the
ejecta were approximately round and, with time, departed more
significantly from the spherical symmetry.\\
This interpretation requires some caution, however, as
\citeauthor{2006PASP..118..722C} observed data that corresponded to a
dominant axis four days prior to our first observation, which we do
not see in our data at the first epoch.  This may reflect the low
$S/N$ of our first observation, rather than an ``oscillating''
evolution in shape of the structure
of the ejecta; with the ejecta appearing approximately spherical at
maximum but departing from a spherical symmetry both before and after
maximum light.

\subsubsection{The data on the dominant and orthogonal axes}
\label{sec:res:domort}
\begin{figure}
\includegraphics[width=7.5cm, angle=270]{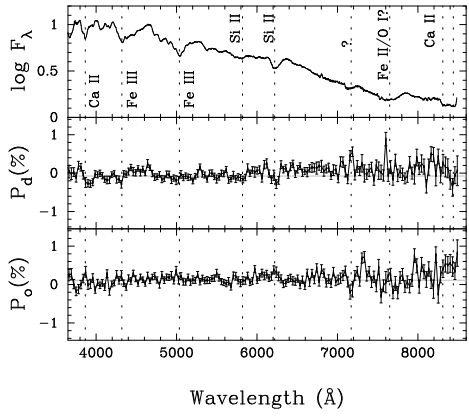}
\caption{Spectropolarimetry of SN~2005hk on 2005 Nov. 9, projected
  onto the dominant axis ($P_{d}$) and the orthogonal axis ($P_{o}$).
  The data have been binned to 30$\mathrm{\AA}$ for clarity, and have
  been corrected for the ISP component.  The weighted average of the
  ISP, rotated onto the dominant and orthogonal axes, is indicated by the heavy grey lines.}
\label{fig:obs:doma}
\end{figure}

\begin{figure}
\includegraphics[width=7.5cm, angle=270]{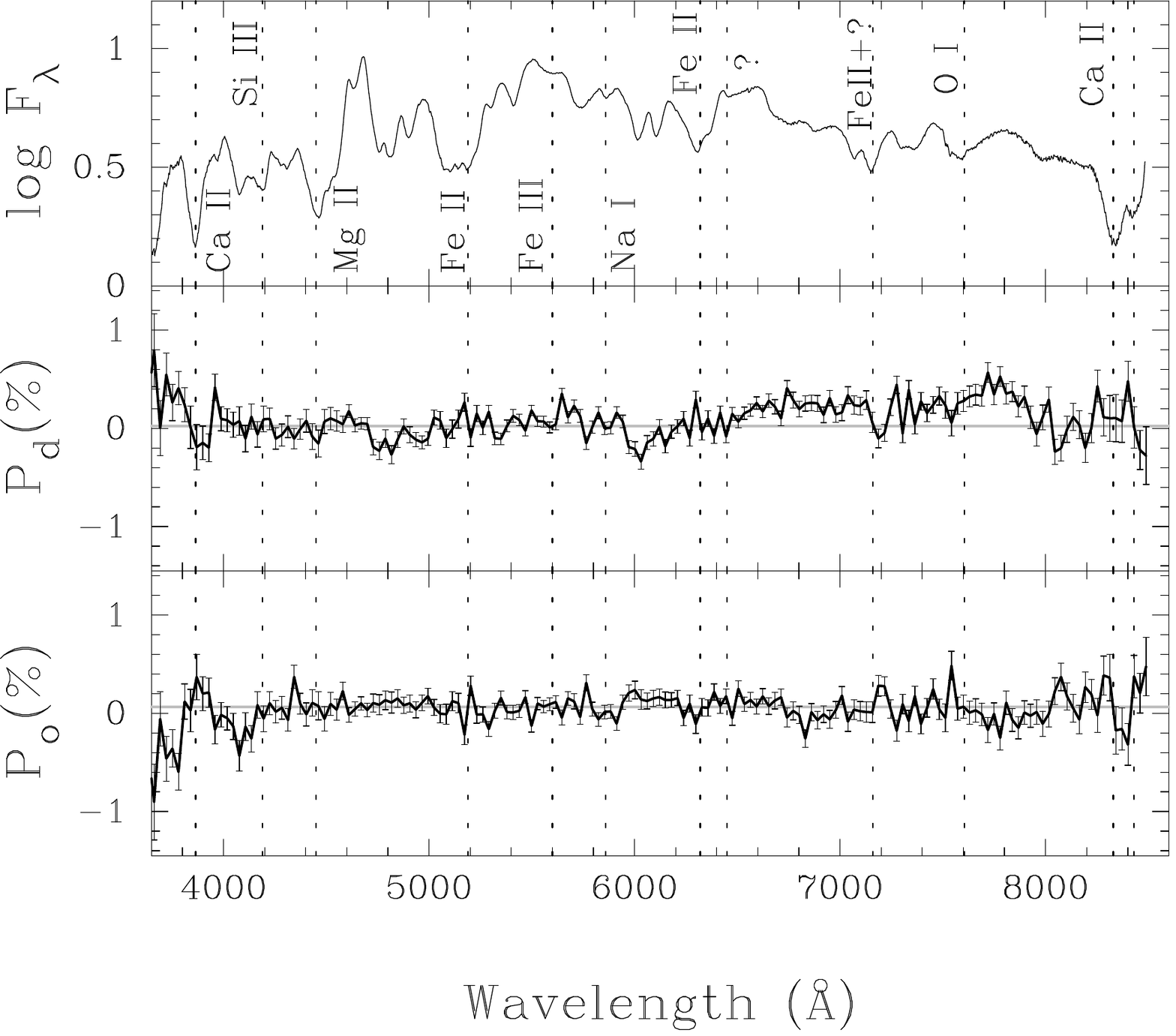}
\caption{Same as Fig. \ref{fig:obs:doma}, but for spectropolarimetry of SN~2005hk on 2005 Nov. 23.}
\label{fig:obs:domb}
\end{figure}

The data were rotated onto the dominant and orthogonal axes, using the
angles determined from the least-squares fit analysis (for the limited
wavelength range of 4000-7000\AA).  The rotated data are presented as
Figs. \ref{fig:obs:doma} and \ref{fig:obs:domb}.  Despite the apparent
increase in the polarization angle from the first to second epochs,
the determined polarization angles for the dominant axes are different
from the angle measured at -4 days, for which
\citeauthor{2006PASP..118..722C} determined
$\theta_{dom}=35.6\degr$.\\
At the first epoch, the dominant and orthogonal polarization
components are similar to Stokes $Q$ and $U$, due to the small angle
through which the data were rotated.  The polarization modulations
associated with \ion{Ca}{2}, \ion{Fe}{3}, \ion{Si}{2} are
predominantly 
observed in the dominant polarization component, and over the range
4000-7000\AA\ the deviations observed in the orthogonal polarization
spectrum are consistent with the expected noise level.  Similar to the
\citeauthor{2006PASP..118..722C} observations at -4 days, an important
exception is the 7200\AA\ absorption feature (and possibly \ion{Fe}{3}
$\lambda 5129$) which has both dominant and orthogonal polarization
components (although \citeauthor{2006PASP..118..722C} refer to the
rotated Stokes parameters $q_{RSP}$ and $u_{RSP}$, these are
equivalent to our dominant and orthogonal axes).  At our second epoch,
the majority of features, associated with \ion{Fe}{2}, \ion{Na}{1}
and, particularly, the $7200\AA$ feature show some
modulation in both the dominant and orthogonal polarization components.  While
\citeauthor{2006PASP..118..722C} observed an obvious wavelength
dependence in the dominant polarization component, with increasing
$P_{d}$ at redder wavelengths (which we estimate to be $0.1\%\
\mathrm{per}\ 1000\AA$), 4 days later we found the wavelength
dependence to be shallower $0.05\%\pm 0.01\ \mathrm{per}\ 1000\AA$.
At $+14$\ days we measure a similar wavelength dependence $0.07\%\pm
0.01\ \mathrm{per}\ 1000\AA$.  This wavelength dependence is clearly
seen on Fig. \ref{fig:obs:qupanelb}.  The change may not be evidence
for evolution in the wavelength dependence of the dominant
polarization component, in itself, as the observations of
\citeauthor{2006PASP..118..722C} were acquired with higher $S/N$.
Such wavelength dependence of the polarization, increasing at redder
wavelengths was observed for SN 1999by \citep{2001ApJ...556..302H}.
At the temperatures determined for SN~2005hk, and for Fe-rich matter,
the opacity at the photosphere is dominated by blends of line transitions for
$\lambda < 6000\AA$ \citep[see Fig. 2 of][]{1993A&A...268..570H}.  At
redder wavelengths, the opacity is primarily due to electron
scattering.   The continuum polarization may indicate an aspherical
Thomson scattering region, e.g. by a global asymmetry induced by the
rotation of the White Dwarf progenitor \citep{2001ApJ...556..302H} or
off-center delayed detonations \citep{2006NewAR..50..470H}.
Because multiple line scatterings are effectively
depolarizing, the observation of increasing polarization with
wavelength may be understood in terms of the decrease of the role of line opacity
at redder wavelengths. If the dominant axis corresponds to a principal axial
symmetry, then the line forming regions of the species observed in the
flux spectrum follow the same axial symmetry as that of the continuum
at the first epoch, but not at the second epoch.

\subsection{The Inferred Ejecta Geometry}
\label{sec:disc:ejecta}
\begin{figure}
\includegraphics[width=6cm, angle=270]{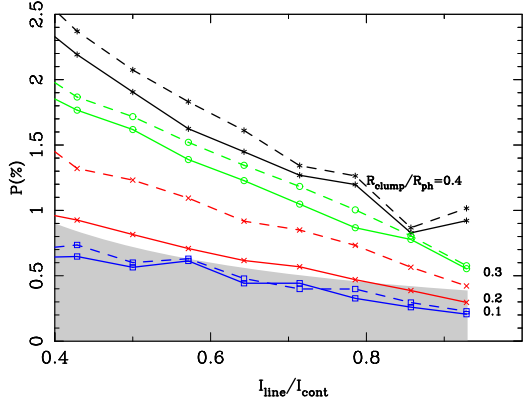}
\caption{Results of Monte-Carlo simulations of random obscuring
  spherical clumps in front of a spherical photosphere.  The models
  are shown for clump scales of $R_{clump}/R_{ph} = 0.1$ (blue
  $\Box$), $0.2$ (red $\times$), $0.3$ (green $\circ$) and $0.4$
  (black $\ast$).  Solid lines are for simulations with $k=0.5$, while
  dashed lines indicate simulations with $k=0$. The shaded area indicates the permitted values of
  line polarization, according to Eqn. \ref{eqn:res:linepollim}, for a
  continuum polarization of $0.36\%$ determined by
  \citet{2006PASP..118..722C}.  Lower values of the continuum
  polarization ($0.17\%$\ determined here) place even stricter upper limits on
  the scale of clumping.  Minor fluctuations in the curves
  are due to the statistical nature of the Monte Carlo simulations; except for the case of
  $R_{clump}/R_{ph}=0.4$ for $I_{line}/I_{cont}>0.8$, where the
  results are dominated by the finite size of the ``pixels'' that compose the model photosphere.}
\label{fig:obs:monte}
\end{figure}

On the basis of the continuum polarizations measured at 0 and 14 days,
any asymmetry in the shape of the photosphere (assuming a spheroidal
configuration) requires an axial ratio $>0.9$
\citep{1991A&A...246..481H}.   A similar limit on the asphericity of
the photosphere is achieved if one uses the higher value of continuum
polarization ($0.36\%$) determined by
\citeauthor{2006PASP..118..722C}\\
The general level of line polarization is low, suggesting that the
line forming regions cover an approximately spherically symmetric
photosphere homogeneously.   Limits on the amount of polarization
expected to be observed can be calculated, given a level of continuum polarization and
line depth, under the assumption that the line forming region blocks only
{\it unpolarized light} from the photosphere, while not blocking
polarized light arising from the limb (i.e. that the photosphere can be
separated into discrete polarized and unpolarized zones):
\begin{equation}
p_{line} \leq p_{cont} . \frac{I_{cont}}{I_{line}}
\label{eqn:res:linepollim}
\end{equation}
where $I_{cont}$ and $I_{line}$ are the observed fluxes in the
continuum and in the absorption line minimum, respectively (i.e. such
that the line depth is $I_{cont}-I_{line}$ - this convention is
chosen as the observed polarization arises from the light that is not absorbed).  As \citet{2006PASP..118..722C} calculated for their observation of SN~2005hk, we also calculate the theoretical limits on the line polarization for measured absorption line depths in our observations. \\
At the first epoch, the maximum polarization for the \ion{Si}{2}
$\lambda 6355$ line is only 0.14\%.  As
\citeauthor{2006PASP..118..722C} suggest, that the lines are not
strongly polarized is commensurate with their apparent weakness.  This
is true at the first epoch, where line strengths of only $>0.7$ are
observed.  At the second epoch, however, absorption line depths of
$\sim 0.5$ are observed, and with the increase in depth of these
lines it might be expected that, should the line forming regions be
asymmetric, that there would also be a commensurate increase in the
polarization observed across these lines (see Fig. \ref{fig:obs:monte}).\\  
Eqn. \ref{eqn:res:linepollim} assumes that all of the polarization is due to the continuum, which implicitly assumes that the photosphere itself is asymmetric.
In a number of cases, in particular for core-collapse SNe, the
classical limit for line polarization is exceeded \citep[see
e.g.][]{maund05bf,2009ApJ...705.1139M}, for low values of continuum
polarization which suggest approximately spherical photospheres.
Violations of this criterion can be explained by large scale structure,
with a dissimilar geometry to that of the photosphere, or clumping,
which may produce polarization irrespective of the shape of the
photosphere or the degree of continuum polarization.  The degree of
observed polarization, for a clumpy line forming region, should depend
on the size of the clumps: for a given line absorption depth, fewer larger clumps block larger areas
with contiguous polarization vectors, whereas more smaller clumps can
be distributed more evenly across the photosphere.\\
In order to explore the possible role of clumping, and the expectation
for the production of polarization across shallow lines, a Monte-Carlo
toy model of a simple photosphere and obscuring clumped line forming
region was constructed following the scheme of
\citet{2007Sci...315..212W}.  A circular continuum emitting
photosphere was adopted, with the luminosity as a function of
projected radius $r_{p}$ across a photosphere of radius $R_{ph}$ set to:
\begin{equation} 
I(r_{p})/I(0)= 1 - k \left( 1 -
    \sqrt{1-\left(r_{p}/R_{ph}\right)^{2}} \right)
\end{equation}
with limb darkening modelled using the coefficient $k$, which for
$k=0$ yields no limb darkening.  For a three-dimensional spherical
photosphere, this would correspond to to a limb darkening law of
$I(r_{p}=R_{ph}\sin\theta)/I(0)=\left( 1 - k \left( 1 - \cos \theta
  \right)\right)$. The radial dependence of the polarization was
assumed to follow a quadratic form $p \propto (r_{p}/R_{ph})^{2}$,
with the constant of proportionality chosen to yield 15\% polarization
for light emitted at the limb of the photosphere
\citep{Chandrasekhar}.  Opaque circular clumps were assumed, with
$\tau >> 1$, such that all photospheric flux from the region
underlying the clump was blocked.  For each iteration of the model,
$N$ clumps of size $R_{clump}/R_{ph}$ were randomly positioned across
the photosphere and the corresponding flux of the uncovered regions,
and their polarization components, were summed.  Clumps were permitted
to overlap and only partially cover the photosphere.  For each size of clumps, 3000 simulations were conducted.
This permits a comparison of the expected polarization for a given
absorption line depth, assuming a particular clump size in the line
forming region.  The results of this model (for $R_{clump}/R_{ph} =
0.1$, $0.2$, $0.3$
and $0.4$) are shown as Fig. \ref{fig:obs:monte}.\\
For some of the models, zero polarization was achieved for uniform
coverage of the photosphere by the clumps, however the difference
between the average polarizations expected (as shown in
Fig. \ref{fig:obs:monte}) and null polarization was $>1\sigma$ for all
the calculated models.  These models suggest that, for a clumpy line
forming region, the clump scale for SN~2005hk is very small ($\leq
0.2R_{ph}$) to produce lines of the depths observed, in the flux
spectra at both epochs, with low polarization.  The uniformly low
polarization observed for lines arising from different species suggest
that the ejecta are generally mixed, with no separate structures {\it
  on the plane of the sky} for different elements or ionization
stages(e.g \ion{Fe}{2} and \ion{Fe}{3}).  The element responsible for
the absorption at 7200\AA\ may be an exception, with a line
forming region that is more aspherical than other species.  The data for
SN~2005hk, at the two epochs presented here and those presented by
\citet{2006PASP..118..722C}, are consistent with an approximately
spherical symmetric photosphere surrounded by similar line forming
regions for all species.\\
The principal assumption of the model presented above is that the line
forming region and the photosphere are separate.  As discussed in
\S\ref{sec:res:domort}, the opacity at the photosphere (where the
quasi-continuum forms) is not due to electron scattering alone, but
rather includes a significant wavelength dependent contribution from
atomic transitions; this implies that line forming regions are not, in
reality, separate entities except in the most high-velocity regions
such as those discussed by \citet{2003ApJ...593..788K}.\\
The size of the clumps and their relation to
"large-scale" structure within the ejecta, leads to an interesting
problem of semantics: since very large clumps ($R_{clump}/R_{ph}=0.4$)
may be considered large scale structure and very small clumps
($R_{clump}/R_{ph}=0.1$) tend towards constituting an effectively
homogeneous medium.  Large opacities require that a large fraction of
the photosphere, the last Thomson scattering distance, are reprocessed
by lines, mostly due to iron-group elements (which will result in a
significantly altered spectrum).

\section{Discussion}
\label{sec:disc}

\subsection{SN 2005hk as a Type Ia SN}
\label{sec:disc:typeia}
\begin{figure}
\includegraphics[width=8.5cm, angle=270]{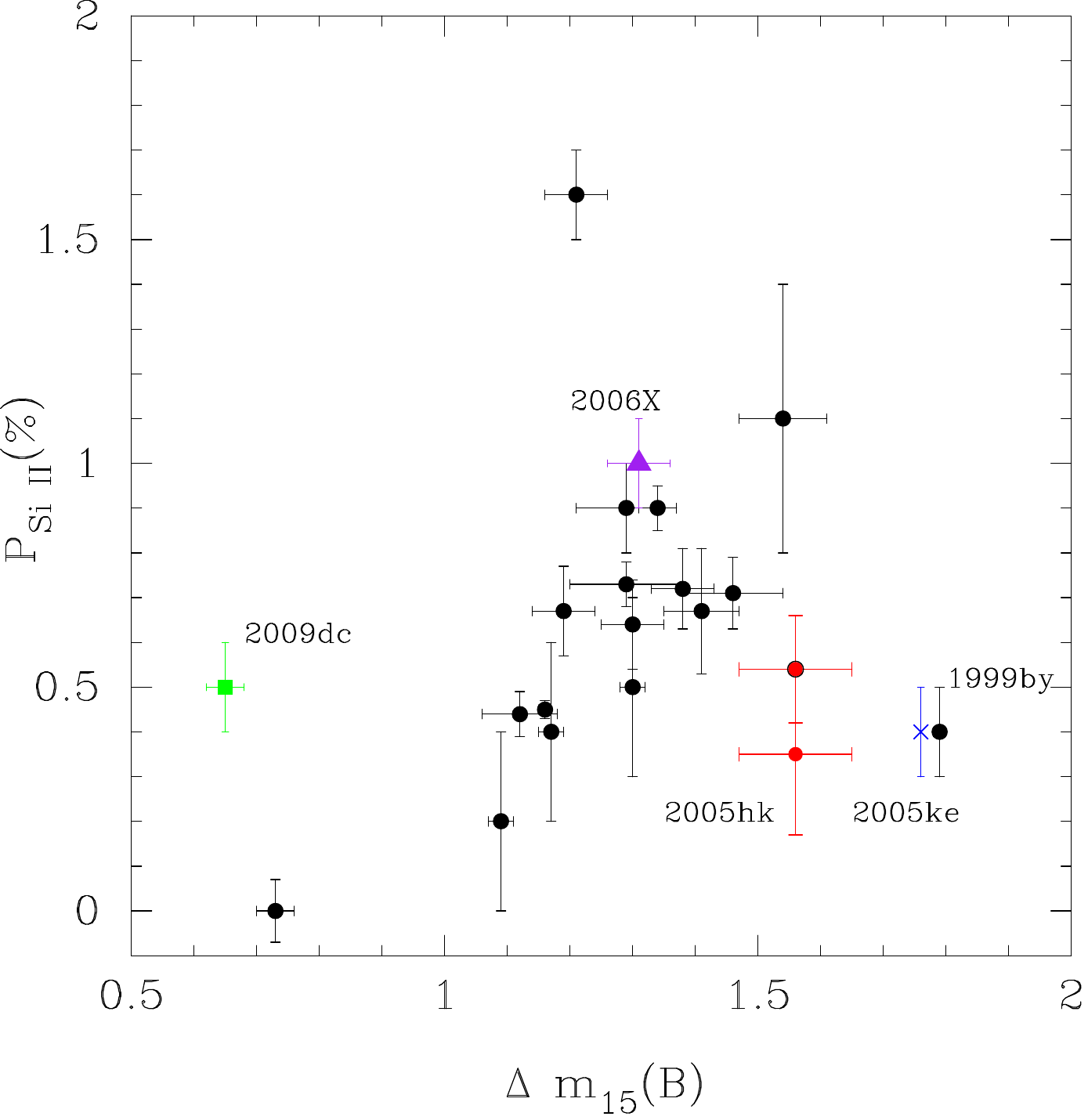}
\caption{Degree of polarization across  Si II $\lambda6355$  as a
  function of light curve decline rate.  Our measurements of SN~2005hk are
  shown in red, with the value of $\Delta P$(\ion{Si}{2}) indicated by
  the point with the heavy outline, data in black are taken from
  \citet{2007Sci...315..212W}, the measurement for SN~2006X
  ($\bigtriangleup$) is from \citet{2009A&A...508..229P}, the
  measurement for the luminous event SN~2009dc ($\Box$) is taken from
  \citet{2009arXiv0908.2057T}, and the measurement for SN~2005ke
  ($\times$) is taken from Patat et al., (2010, in prep.).}
\label{fig:disc:wang}
\end{figure}

The class of SNe similar to SN~2002cx currently has 15 members,
including 2005hk and 2008ha \citep{2009AJ....138..376F}.  As discussed
in detail by \citet{2003PASP..115..453L} and
\citet{2007PASP..119..360P}, both SN~2002cx and SN~2005hk were of
especially low velocity at the photosphere near maximum light, $\sim
6000$ \kms, had early relatively high ionization, and low peak
luminosity.  Neither showed the typical secondary maxima in R and I of
normal SN~Ia, but were also dissimilar to previously studied
under-luminous SN~Ia that do not show the secondary maxima \citep[e.g.][]{1992AJ....104.1543F}.  There is
also extreme diversity within the class, with SN~2008ha being $\approx
3$ magnitudes fainter than SN~2002cx \citep{2009AJ....138..376F}.  The
light curve decline is rather slow, unlike previous under-luminous
SN~Ia that serve to define the light curve brightness/decline
relation prominently used in cosmology studies.\\
\citet{2003PASP..115..453L} pointed out that no published model of a
SN~Ia corresponds to all the behavior of SN~2002cx, but suggested that
a pulsating delayed detonation \citep{1991A&A...245..114K} model might
be worth considering further. \citet{2006AJ....132..189J} suggested
that the nebular spectral characteristics that they determined might
be consistent with a pure deflagration model since the models in the
literature have insufficient energy to correspond to normal SNe~Ia and
hence tend to have smaller expansion velocities and produce less
$^{56}$Ni and hence tend to be dim. \citet{2007PASP..119..360P}
suggest that the low expansion velocities, low peak luminosity, and
the absence of secondary maxima in the NIR light curves may be
reproduced by a 3D pure-deflagration model
\citep{2004PhRvL..92u1102G,2006A&A...453..203R} that produces
0.25$M_{\odot}$ of $^{56}$Ni.\\
A key issue with the hypothesis that this class of explosions
represents pure deflagrations of white dwarfs is that the pure
deflagration models leave a substantial amount of unburned C and O in
the outer layers. This C and O should show up in the early phases and while
\ion{O}{1} $\lambda 7774$ is clearly present at the second epoch (and
possibly the first), the case for \ion{C}{2} $\lambda 6578$  (and possibly
for a contribution to the $7200\AA$ feature) is less clear but it is not
conclusively absent. In addition, \citet{2005A&A...437..983K} suggest that the late-time
spectra of pure deflagration models show no similarity to that of
SN~2002cx either, in particular the absence of strong features due to C and
O.  \citet{2006AJ....132..189J} claim a marginal detection of
\ion{O}{1} in late-time nebular spectra of SN~2002cx, with the
suggestion that more unburned C and O may be contained inside the
dense, still optically thick core at late times.  
\citeauthor{2008ApJ...680..580S} claim that at early times \ion{O}{1}
$\lambda 7774$ is suppressed by high temperatures and, at late times,
nebular \ion{O}{1} lines are also suppressed by high temperatures.
The lack of observation of \ion{O}{1} $\lambda 7774$ for SN~2005hk at
very early times (before our observations) is uncertain, due to poor spectral coverage at early times
\citep{2007PASP..119..360P} and low S/N \citep{2008ApJ...680..580S}.
\citet{2008ApJ...680..580S} are able to approximately model the
photospheric spectrum of SN~2005hk, based on the W7 deflagration
explosion model \citep{1984ApJ...286..644N}.  Importantly,
\citeauthor{2008ApJ...680..580S} find that, given the conditions of
the SN, their spectral synthesis suggests large amounts of unburned O.
Their model also assumes spherical symmetry, for which they achieve a
reasonable approximation to the line profiles (which are somewhat
sensitive to any departures from spherical symmetry; see
e.g. \citealt{2008arXiv0807.1674T} for a counter example).  The
features remain too broad, are forbidden rather than permitted, and
show prominent emission of [\ion{O}{1}] $\lambda 6300,6364$. High
densities, as evidenced by the low expansion velocities and P Cygni
profiles observed at late nebular epochs, might suppress the
[\ion{O}{1}] feature, but it is predicted in the models and so its
absence is a definite constraint on the nature of SN~2005hk.\\
\citet{2006PASP..118..722C} consider the possibility that these events
may be related to normal SNe~Ia, but viewed along the ``hole" left by
having the binary companion nearly directly along the line of sight
\citep{2000ApJS..128..615M,2004ApJ...610..876K}. Such an aspect might
increase the prominence of high-ionization species, weaken absorption
lines, reduce the photospheric velocity and induce some polarization
due to the non-spherical geometry of the ejecta. One issue with this
model for SN~2002cx and related explosions is that it would result in
higher luminosities than when viewed from other aspect angles. Even if
the underlying explosion were a subluminous event such as SN~1991bg,
if the view were ``down the hole," the result would be brighter than
SN~1991bg. Both SN~2002cx and SN~2005hk were subluminous by about 1
magnitude in all bands \citep[except H;][]{2007PASP..119..360P}.  It
is certainly not clear that this is consistent with a ``hole"
model. \citealt{2006PASP..118..722C} comment
that their measurement of a small but appreciable continuum
polarization and weak line features was inconsistent with the
prediction of \citet{2004ApJ...610..876K} of a continuum
polarization of $\sim$ 0.1 percent and strong line polarization peaks
for an explosion viewed slightly off the axis of the hole.  As the SN
evolves, the relative size of the hole and the photosphere will change
(in particular if the hole closes, which \citealt{2004ApJ...610..876K}
expect to be a slow process if it occurs), which should lead to a
commensurate change in the polarization properties.  Importantly, we
see only moderate change in the polarization, despite the significant changes
in the flux spectrum, between the epochs at which SN~2005hk was
observed.  In addition, it is not at all clear how such a model would
account for the late-time permitted line spectrum delineated by
\citet{2006AJ....132..189J}.  Our observations do suggest some line
polarization. This demands some asymmetry in the ejecta, consistent
with a single axial symmetry, but
does not spell out quantitative agreement with the ``hole" model.\\
An alternative model might be a ``plume" model
\citep{2004ApJ...612L..37P,2005ApJ...632..443L} in which a single
plume (or very few plumes) reach the surface, spreading fresh
$^{56}$Ni.  If the plume burned little, the expansion might be slow
and leave high densities at late times.  Our spectropolarimetry
suggests that the outer layers of the ejecta are approximately
homogeneously mixed, although stratified, which suggests mixing in a
large number of
smaller plumes (with size $\leq 0.2R_{ph}$ at the first epoch).\\
This class of objects thus presents many conundra. They show high
excitation early on but otherwise have none of the properties of SNe~Ia
like SN~1991T that show early \ion{Fe}{3}. Their light curves are dim
but broad, violating the ``Phillips relation''. They have very slow
expansion velocities and may remain partially optically thick at
nearly 300 days after the explosion. They are mildly asymmetric in the
continuum, in the \ion{Fe}{3} early on and in \ion{Si}{2} near
maximum.  The polarization of SN~2005hk is completely compatible with
the observed range of polarization properties for Type Ia SNe (see
Table \ref{tab:disc:comp}).  The level of polarization of SN~2005hk is
not the lowest {\it detected} for a SN, with SN 1996X (a normal Type Ia SN;
\citealt{1997ApJ...476L..27W}) showing polarization just above
statistical significance and with no apparent dominant axis.  The
observations of SN~2005hk show it to have a spectropolarimetric type
of D1 \citep{2008ARA&A..46..433W}, similar to most Type Ia SNe, indicating that there is a
principal axis of symmetry to the ejecta.  Importantly, there are no
apparent loop like features, even for the strongest lines (which is
particularly pertinent for the observations at the second epoch) which
would indicate a significant departure from this main axial symmetry.
Normal Type Ia SNe, such as SN 2001el
\citep{2003ApJ...591.1110W} and SN~2004S \citep{2006astro.ph..9405C},
show some structure in their ejecta principally in the \ion{Si}{2}, as
well as the high velocity component of the \ion{Ca}{2} IR3 \citep[see
also][]{2003ApJ...591.1110W,2009A&A...508..229P}.  Conversely, there
are Type Ia SNe, such as SN~2004dt \citep{2006ApJ...653..490W}, whose
line polarization is principally in the dominant polarization
component which is consistent with a single axial symmetry.  In the
case of SN~2004dt, however, the observed line polarization was $\sim
1\%$.  In summary, the similarity between the spectropolarimetry of
SN~2005hk and Type Ia SNe indicates that their respective explosion
mechanisms are both approximately spherical (see Table \ref{tab:disc:comp}).\\
\citet{2007Sci...315..212W} identified a correlation between the
degree of polarization of \ion{Si}{2}, at -5 days, with the decline
rate ($\Delta m_{15}(B)$) of the light curve.  At some wavelengths
SN~2005hk is an outlier from the Phillips brightness-decline
relationship, although it is consistent with the relationship in the $I$
and $H$ bands \citep{2007PASP..119..360P}.  While the polarization measurements of
SN~2005hk is offset from the measured relation derived for normal
Type~Ia SNe by \citeauthor{2007Sci...315..212W}, it is not as significant an outlier as the subluminous
Type~Ia SNe 1999by and 2005ke (see Fig. \ref{fig:disc:wang}).\\

\subsection{SN 2005hk as a Core-Collapse SN}
\label{sec:disc:ccsn}
\citet{2006AJ....132..189J}, \citet{2009Natur.459..674V} and
\citet{2009AJ....138..376F} consider the possibility that events such
as SN~2002cx and 2005hk are not related to Type Ia SNe, but may
instead be a type of CCSN.  Based on the photometric and spectroscopic
similarity between SN~2008ha and SN~2005hk with low-luminosity Type
IIP and Type Ic SNe (2005cs and 2007gr, respectively) and the
occurrence of these events in late-type galaxies,
\citet{2009Natur.459..674V} suggest 2002cx-like SNe are low-luminosity
Type Ic SNe, potentially associated with Gamma Ray Bursts (GRBs).
\citet{2009AJ....138..376F} present additional events in this class,
which are claimed to have taken place in lenticular S0 galaxies, which
may argue against massive stars as the universal progenitors for this
type of event.  Spectropolarimetry provides additional parameter space
in which to examine the similarities between different types of SNe.\\
\citet{2009Natur.459..674V} and \citet{2009AJ....138..376F} note the
spectroscopic similarity between SN~2005hk and SN~2008ha at +15 and +8
days (after optical maximum), respectively, as well as their similarity with the subluminous
Type IIP SN 2005cs at +44 days, whilst it was still on the photometric
plateau \citep{andrea05cs}.  At such times Type IIP SNe have generally
low polarizations \citep{2006Natur.440..505L}, seemingly similar to
that observed by us for SN~2005hk. From a spectropolarimetric
perspective, however, comparison of SN~2005hk with SN~2005cs whilst on
the plateau is inappropriate.  During the plateau phase, in Type IIP
SNe, the photosphere is located in the hydrogen envelope.  Due to the
low ionization energy of H, the shape of the photosphere is
approximately spherical.  Once the photosphere has descended below the
H envelope, into the He layer (He being harder to ionize and excite),
the shape of the photosphere more closely follows the shape of the
underlying exciting material.  Despite an asymmetric explosion,
\citet{2006Natur.440..505L} observed that during the plateau SN~2004dj
had a low polarization, which increased significantly once the plateau
phase had ended.  Given the H-deficient nature of SN~2002cx-like SNe,
a more appropriate comparison is with the H-deficient CCSNe, which
have been observed (as listed in Table \ref{tab:disc:comp}) to have
significant continuum and line polarization for different species
indicating the role of highly asymmetric explosion mechanisms.  Highly
asymmetric interiors have also been observed for the Type IIP SNe
2001dh and 2003gd (Maund et al., 2010, in prep.), which suggest that
an asymmetric explosion mechanism is common to all these CCSN events. \\ 
It is interesting to note another ongoing controversy
involving SNe~Ia and SNe~Ic. SN~2002ic was revealed by
\citet{2003Natur.424..651H} to resemble a SN~Ia embedded in a
hydrogen-rich medium. \citeauthor{2003Natur.424..651H} interpreted this
as a SN~Ia exploding with an evolved binary companion.
\citet{2004ApJ...604L..53W} showed that the hydrogen component was
strongly polarized and argued that the geometry was consistent with an
extended, clumpy, disk-like structure such as observed in
proto-planetary nebula stars.  \citeauthor{2003Natur.424..651H} and
\citet{2004ApJ...604L..53W} identified other events of this category
that were originally identified as Type IIn events, but which, on
closer inspection and comparison with SN~2002ic, revealed an
underlying Type I spectrum. More recently, \citet{2006ApJ...653L.129B}
have challenged this interpretation, arguing that the underlying
supernova in SN~2002ic more closely matched a SN~Ic than a SN~Ia.\\ 
In summary, the nature of the polarization of SN~2005hk is
inconsistent with the polarization measured for CCSNe, in particular
those of Type Ibc which have been generally observed to have very
large polarization (especially line polarization) that increases with time.

\section{Conclusions}
\label{sec:conc}
SN~2005hk has been observed at two epochs, corresponding to maximum
light and two weeks after maximum.  After subtraction of the ISP
component, the SN is observed to have a very low to null intrinsic
polarization.  A modulation in the Stokes parameters is observed across the
\ion{Si}{2} $\lambda 6355$ line being the most significant feature, at
the first epoch, with $\Delta P = 0.54\pm0.12\%$.\\
At both epochs, a dominant axis can be fit to the data, although the
data are poorly described by simple straight line fits on the Stokes
$Q-U$ plane.  While there is modulation in the dominant polarization
component, at the first epoch, that is associated with line features in
the flux spectrum, the absence of similar modulations in the
orthogonal polarization component implies that the line forming
regions of the species in the flux spectrum share the same axial
symmetry as that of the photosphere.  The data at both epochs is
classified as being of Spectropolarimetric Type D1.\\
In comparison with previously published high S/N observations of
SN~2005hk at -4 days \citep{2006PASP..118..722C}, we
observe a possible rotation of the polarization angle of the dominant
axis, but it is unclear (due to the differences in S/N) if this
represents a real rotation of the principal axis of symmetry with depth
into the ejecta.\\
The low level of continuum polarization limits the asymmetry of the
photosphere, at both epochs, to $<10\%$.  The low line polarization is
suggestive that the line forming regions are mixed, that different
species occupy similar locations within the ejecta, and that they have
a similar geometry to that of the photosphere.  This is similar to the
general behaviour of Type Ia SNe, and different from the chemically
segregated ejecta observed for
CCSNe.\\
We conclude that the spectropolarimetry of SN~2005hk is not
inconsistent with some of the expectations of the deflagration
explosion model.  We find that, in order to preserve the
approximate spherical symmetry of the photosphere and the line forming regions,
any plume-like structures would be required to have a scale, in the
plane of the sky, of $\leq 0.2R_{ph}$ at both epochs.  This would give
the overall appearance of homogeneously mixed ejecta containing both
the products of nucleosynthesis (Fe, Ni, Co) and unburned O at
all depths.\\
Spectral similarities between SN~2002cx-like events and low-luminosity
Type IIP SNe are likely to be merely superficial. The differences in
the observed spectropolarimetry of SN~2005hk with CCSNe suggests, from
a geometry perspective, that these events have different underlying
explosion mechanisms.  Unfortunately, with spectropolarimetry of only
one SN from the 2002cx-like class it is not possible to determine if
these results are applicable to the entire class, in particular to the
most peculiar member of this class, SN~2008ha.
\section*{Acknowledgements}
The research of JRM is funded through the Sophie \& Tycho Brahe
Fellowship.  The Dark Cosmology Centre is supported by the DNRF.  The
research of JCW is supported in part by NSF grant AST-0707769.  PAH is
supported by the NSF grants AST 0708855  and 0703902.  The
authors are grateful to the European Organisation for Astronomical
Research in the Southern Hemisphere for the generous allocation of
observing time. They especially thank the staff of the Paranal
Observatory for their competent and never-tiring support of this
project in service mode.  The authors thank the anonymous referee for
useful comments and suggestions that have improved the manuscript.

\bibliographystyle{apj}

\begin{table}
\caption{\label{tab:obs:log}Spectropolarimetry of SN~2005hk}
\begin{tabular}{ccccc}
\hline\hline
Object   &  Date &  Exposure &  Median  & Type  \\
         &  UT   &   (s)     &  Airmass &       \\
\hline
SN~2005hk&2005 Nov 09.163 &$4 \times 900$ & 1.255 & SN \\ 
HD~49798 &2005 Nov 09.363 & 5             & 1.067 & Flux Std.\\
\\
SN~2005hk&2005 Nov 23.096 &$4 \times 900$ & 1.160 & SN \\ 
LTT~1020 &2005 Nov 23.132 & 5             & 1.139 & Flux Std.\\
\hline\hline
\end{tabular}
\end{table}
\begin{table}
\caption{\label{tab:res:delp} Polarization changes across spectral features}
\begin{tabular}{lccc}
\hline\hline
Species & $\lambda_{cont}(\AA)$ & $\lambda_{line}(\AA)$ & $\Delta P$\\
\hline
\multicolumn{4}{c}{{\bf 2005 Nov 09}}\\
\ion{Ca}{2} H \amp K & 3780 & 3960 & $0.64\pm0.15\%$\\
\ion{Fe}{2} & 4225 & 4315 & $0.34\pm0.11\%$\\
\ion{Si}{2} & 6120 & 6270 & $0.54\pm0.12\%$\\
\ion{Fe}{2}? & 7070 & 7215 & $0.45\pm0.17\%$ \\
\ion{Ca}{2} IR3 & 8340 & 8490 &$0.31\pm0.40\%$\\
\hline
\multicolumn{4}{c}{{\bf 2005 Nov 23}}\\
\ion{Ca}{2}  H \amp K & 3750 & 3930 & $0.49\pm0.23\%$\\
\ion{Fe}{2} & 4017 & 4077 & $0.54\pm0.19\%$ \\
\ion{Fe}{2} & 4610 & 4910 & $0.35\pm0.12\%$\\
\ion{Fe}{2} & 5910 & 6030 & $0.5\pm0.11\%$\\
\ion{Fe}{2}? & 7130 & 7220 & $0.34\pm0.14\%$ \\
\hline
\end{tabular}
\\
\end{table}
\begin{table*}{\small
\caption{\label{tab:obs:domaxis} Best-fit Dominant Axes}
\begin{tabular}{cccccccccccc}
\hline\hline
                      & &\multicolumn{4}{c}{Least Squares Fit
                        Analysis} & & \multicolumn{2}{c}{Principal
                        Components} & & \multicolumn{2}{c}{Centroids}
                      \\
\cline{3-6}\cline{8-9}\cline{11-12}
Date             & & $\alpha$        & $\beta$        & $\chi^2/dof$ & $\theta_{dom}$ & & Axial Ratio & $\theta_{e}/2$ & & $\overline{Q}$  & $\overline{U}$    \\
                 & &                 &                &              &              & &$b/a$         &                & &               &                  \\
\cline{3-6}\cline{8-9}\cline{11-12}
09 Nov 2005$^{A}$& & $0.138\pm0.008$ & $0.077\pm0.074$ & 347.2/154   &$2.2\pm7.7$   & &  0.87        & 3.7             & & $-0.037\pm0.016$&$0.135\pm0.016$  \\
09 Nov 2005$^{B}$& & $0.138\pm0.009$ & $0.226\pm0.084$ & 202.3/98    &$6.4\pm7.9$   & &  0.78        & 8.7              & & $-0.049\pm0.016$&$0.127\pm0.016$   \\
\\
23 Nov 2005$^{A}$& & $0.075\pm0.008$ & $0.545\pm0.051$ & 288.8/154   &$14.3\pm5.0$  & & 0.47         & 14.0             & & $0.011\pm0.016$&$0.081\pm0.015$\\
23 Nov 2005$^{B}$& & $0.079\pm0.010$ & $0.771\pm0.077$ & 160.9/98    &$18.8\pm5.0$  & & 0.42         & 18.55            & & $-0.022\pm0.016$ &$0.062\pm0.016$        \\
\hline\hline
\multicolumn{11}{l}{$^{A}$ for data over the wavelength range 3700-8600\AA}\\
\multicolumn{11}{l}{$^{B}$ for data over the wavelength range 4000-7000\AA}\\
\end{tabular}
}
\end{table*}
\begin{table}
\begin{centering}
\caption{\label{tab:disc:comp}Comparison SN Spectropolarimetry}
\begin{tabular}{lcccccc}
\hline\hline
SN          & Spectral  & Epoch$^{\dag}$  & SP          &  Continuum  &  Line      & Ref. \\
            & Type      &                & Type$^{\ddag}$  &   Pol. (\%) &  Pol. (\%) &      \\
\hline
1996X      &   Ia (normal) &   -7           &    N1        &  $<0.3$     &  \ion{Si}{2}:$0.6\pm0.15$ &   1   \\  
1999by     &   Ia (sub) &   -1           & D1          & $0.8\pm0.1$ &             &  2    \\
2001el     & Ia (normal) & +2            & D1+L(Ca)    & $0.3$  &Si:0.3;Ca:0.7 & 3   \\ 
2003du     & Ia (normal) &  +18          & D1   &    & $0.2$ & 4 \\ 
2004S      & Ia (normal) & +9            & D           &  $0.4\%$    & $\leq 0.5$ &  5     \\
2004dt     & Ia (normal) & -7            & D1(Si+Mg)   &             & \ion{Si}{2}:$2$  &  6 \\
2006X      & Ia (normal) & 0             & D1+L        & $<0.2$      & \ion{Si}{2}:$0.8$ & 7 \\
\hline\hline
2005hk & 2002cx-like& -4 & D1  & $0.4$ & & 8 \\
$\cdots$ & $\cdots$  & 0  & D1  & $\sim 0.2$& & this work \\
$\cdots$ & $\cdots$  & 14 & D1  & $\sim 0.2$& & this work \\
\hline\hline
2002ap & Ic (broad-lined)    & 0  & DL  & $\sim0.3\%$& \ion{O}{1}:4 & 9 \\
2005bf & Ib (pec)    & -6$^{\ast}$ & D1,L(He+Ca) & $<0.45\%$ & \ion{He}{1}:1.3,\ion{Ca}{2}:4 & 10 \\
2007gr & Ic   & +21 & D1+L(Ca) & $0.5\%$ & \ion{Ca}{2}:2.5 & 11 \\
2008D  & Ib   &0 & D1+L(He+Ca) & $<0.4\%$ & \ion{He}{1}:0.6,\ion{Ca}{2}:1.8 &12\\
$\cdots$&$\cdots$ &14 & D1+L(He+Ca) & $<0.4\%$ & \ion{He}{1}:1.1,\ion{Ca}{2}:2.5 &12\\
\\
\end{tabular}
\end{centering}
\\
$^{\dag}$ Relative to optical maximum; $^{\ddag}$ SN
Spectropolarimetric types see \citet{2008ARA&A..46..433W}; $^{\ast}$
relative to the second maximum; References (1) \citet{1997ApJ...476L..27W}; (2) \citet{2001ApJ...556..302H}; (3) \citet{2003ApJ...591.1110W}; (4) \citet{2005ApJ...632..450L}; (5) \citet{2006astro.ph..9405C}; (6) \citet{2006ApJ...653..490W}; (7) \citet{2009A&A...508..229P}; (8) \citet{2006PASP..118..722C}; (9) \citet{2003ApJ...592..457W}, \citet{2002ApJ...580L..39K}, \citet{2002PASP..114.1333L}; (10) \citet{maund05bf}; (11) \citet{2008arXiv0806.1589T}; (12) \citet{2009ApJ...705.1139M}.\\
\end{table}
\end{document}